\documentclass[aps,pra,twocolumn,preprintnumbers,amsmath,amssymb,floatfix,groupedaddress]{revtex4}

\usepackage{enumerate}
\usepackage{graphicx}
\usepackage{color}

\begin{document}

\title{Extracontextuality and Extravalence in Quantum Mechanics.}

\author{Alexia Auff\`eves$^{(1)}$ and Philippe Grangier$^{(2)}$}

\address{ (1): Institut N\' eel$,\;$BP 166$,\;$25 rue des Martyrs, F38042 Grenoble Cedex 9, France. \\
(2): Laboratoire Charles Fabry, IOGS, CNRS, Universit\'e Paris~Saclay, F91127 Palaiseau, France.}

\begin{abstract}
We develop the point of view where Quantum Mechanics results from the interplay between the quantized number of  ``modalities'' accessible to a quantum system, and the continuum of  ``contexts'' that are required to define these modalities. We point out the specific roles of ``extracontextuality'' and ``extravalence'' of modalities,  and relate them to the Kochen-Specker and Gleason theorems.
\end{abstract}
\maketitle

\section{Introduction and motivation.}

During recent years a great number of convincing  experiments  have vindicated beyond reasonable doubt the violation of Bell's inequalities \cite{B1,B2,B3,B4,AA},  as well as of inequalities derived from the Kochen-Specker theorem \cite{KS}, and of many variants of them \cite{GHZ}.  These violations are most often taken in a negative sense, as no-go results excluding the possibility of local and non-contextual hidden-variable theories. One can thus conclude that hidden-variable theories, if any, must be non-local and contextual; an example is the de Broglie-Bohm theory \cite{laloe}.  

From an empirical perspective,  all the experiments quoted above, as well as many others, are in perfect agreement with quantum predictions, as well as with relativistic causality (or ``no-signalling''), and  no way to access the hypothetical hidden-variables has ever been found.  
Therefore in this paper we will attempt to make one step further, and assume once for all that there are no hidden-variables whatsoever behind quantum mechanics (QM). If this is taken as an accepted fact, then what are the consequences for quantum theory~? 

In order to answer this question, we will follow the route opened in \cite{csm1,csm2}, by first forgetting the quantum formalism, and restarting from physical considerations. This approach will be illustrated by various examples, until we meet again the mathematical description. 

\section{Contexts, Systems, Modalities.}

As a first step, we will consider a minimal set of hypotheses, corresponding to a simple version of the Kochen-Specker theorem \cite{AC}, and which taken together lead to a contradiction. By examining which of these hypotheses might be given up, and how, we will get a new set of hypotheses, escaping the previous contradiction by using ``extracontextuality'', a concept already introduced in ref.  \cite{csm2}. We will show that this new set of hypotheses is consistent with quantum experiments; it may thus be used as a basis to tell what quantum objects really are. 
Again in agreement with previous work  \cite{csm1}, our conclusion will be that a quantum object, able to carry fully predictable and repeatable properties called modalities, must involve both a system and a context.  
\\

Let us now recapitulate qualitatively a few definitions, that have been written down more formally in \cite{csm2,csm1}, and are summarized by the acronym CSM (Contexts, Systems, Modalities).
 \begin{itemize}
\item A {\bf system}  is an entity of the natural world that can be isolated well enough to carry physical properties with definite values, such as mass, position, angular momentum...
\item These physical properties can be measured using apparatus external to the system. The completely specified measurement  device is called a {\bf context}, and is defined within the framework of classical physics, as it is done in any experiment. 
 \end{itemize}
 These definitions of systems and contexts are valid both in classical and quantum physics, but classically  the context can be regarded  as a tool, allowing one to measure physical properties  which are then carried by the system alone. However, this is no more true in quantum physics, where we posit the following axioms, built up from empirical evidence~:
  \begin{itemize}
\item Axiom 1 : In quantum mechanics  physical  properties can be measured repeatedly, and the results can be predicted with certainty, as long as both the system and the context are kept the same. The set of definite (fully predictable) values of the physical properties is called a {\bf modality}, and it belongs jointly  to the system and the context. 
\item Axiom 2 : Within a context, there are $N$  mutually exclusive modalities, where $N$ depends on the system but is the same in all relevant contexts. 
\item Axiom 3 : The (classical) parameters defining the context can be varied continuously and reversibly; they may be e.g. the orientation of a polarizer, or  the position of a detector.
 \end{itemize}
 As a summary, the ``quantum object'' carrying fully predictable physical properties is the indivisible association of a system and a context (Axiom 1), and these physical properties are quantized (Axiom 2) in any context chosen within a continuous range (Axiom 3). 
 
 In the following we will illustrate these ideas using three examples, which all fit within the CSM framework, but illustrate various aspects, related to quantum interferences in section 3, to Bell's inequalities in section 4 and to Kochen-Specker contextuality in section 5. The reader may prefer one of these illustrations or the other, depending on his/her own background.

We note that our objective here  is not to look for more no-go theorems,  but to exploit definitions related to our CSM version of quantum ontology \cite{csm1} and deduce the QM formalism from them \cite{csm2}. So to be clear we will use different words like extracontextuality, extravalence (i.e. equivalence with respect to extracontextuality), keeping in mind that we are working in the CSM ontological framework \cite{csm1}, different from the usual (classical) one.

 \section{Extravalence of modalities in an interferometric toy model.}

We consider first a simple (but idealized) experiment, and impose some requirements on it, 
which would obviously be verified by an actual  set-up,  up to some technical imperfections. 
So let us consider a quantum particle (system) propagating in a set of $N$ transmission lines (context), 
see Fig.~1. For instance, it may be a photon in an array of $N$ parallel optical fibers. 
\begin{figure}[!h]
\centering
\includegraphics[width = 8.5cm]{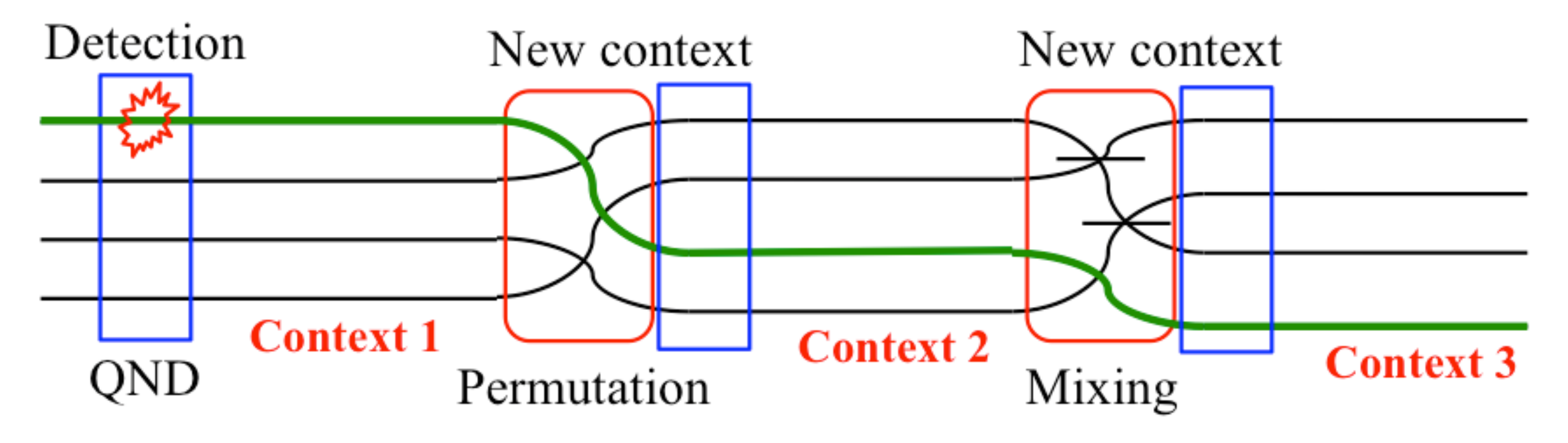}
\caption{Various contexts corresponding to one quantum particle (system) moving along 4 transmissions lines. After changing context (mixing the lines), a new modality is defined by a non-destructive detection (blue box). The different modalities connected by the thick green line are extravalent, i.e. connected with certainty 
through the contexts 1, 2 and 3. }\label{context}
\end{figure}

\noindent {\bf A. Toy model for context, system, modalities. }

In this toy model there is one particle only, i.e. if one looks at all lines, one gets one ``click'' only, which tells in which line the particle is, and all possible clicks are mutually exclusive. This corresponds to the fundamental quantization postulate that we introduced in axiom 2. 

This measurement can be done in a destructive way (by cutting the array of lines and putting a detector on each one), but more interestingly in a non-destructive way, i.e. the click can be obtained without extracting the particle from the array. 
The non-destructive measurement gives always one click, and the line in which  it occurs is certain and repeatable, as long as the context is not changed, in agreement with axiom 1.

The possible results are defined as the modalities corresponding to our system, and are certain, repeatable, and mutually exclusive provided that there is no change in the  context.  A given modality is thus considered to belong jointly to the system (the particle) and to the context (a given array of transmission lines). 
\\

\noindent {\bf B. Extravalence and extracontextuality.}

Then one may change the context by mixing the lines in the array (Fig.1). The most obvious way to do this is by permutation, and then one can follow each modality, and guarantee the certainty by simply rearranging the various modalities (or transmission lines). 

Since the context has (physically) changed after the permutation, we will not say that the modalities are the same (because a modality is associated with a context), but as the certainty  is maintained, one can say that this certainty is transferred from one modality to the other. This is clearly an equivalence relation : certainty is transferred from one modality  to itself (in the case where the context remains the same), and the relation is symmetric and transitive between contexts. This equivalence relation by certainty transfer will be called {\bf  ``extravalence''}.

Though straightforward  in the previous example, the existence of extravalence classes relating modalities in different contexts is by no mean obvious, but it is fully compatible with our postulates. We call this  {\bf  ``extracontextuality''}. 
\\

\noindent {\bf C. Probabilities when changing contexts.}

In general, the change of context is not simply a permutation, but can involve any mixing of all the lines, like on the right side of Fig.~1. According to axiom 2, after this mixing there are always $N$ lines, and the (unique) particle cannot be lost. 
In addition, different modalities belonging to the same extravalence class are connected with certainty (i.e. with probability $p=1$) when changing the context. Therefore the $N$ mutually exclusive modalities in context $C_1$ correspond to $N$ mutually exclusive extravalence classes, and the  same is true for context $C_2$. 

One may thus ask the question: given the modality (line)  $u_i$ in context $C_1$, what modality (line) $v_j$ will be found in context $C_2$ ? 
If a modality $v_j$ can be identified with certainty for any other context $C_2$,  given $u_i$, this may mean different things:

- there is always a deterministic link between the contexts, like for the permutation in Fig.~1. But then one has always the same $N$ mutually exclusive alternatives, and the context can be forgotten. In such a case the modality can be identified with its extravalence class, and attributed to the system alone, like in classical physics. 

- there exists a (possibly hidden) context specifying modalities in both contexts $C_1$ and $C_2$. But such a context would have more than $N$ mutually exclusive modalities, in contradiction with axiom~2. For clarity, let us note that extravalent modalities taken from different contexts cannot be put together to increase the number of modalities within the same context: this would also contradict axiom~2  since only mutually exclusive modalities should appear in the same context. This argument is closely connected to the Kochen-Specker theorem, that will be discussed in the next section. 

So in the general case, contextual quantization requires that given modality $u_i$ in context $C_1$, only the {\bf probability} to get modality $v_j$  in context $C_2$ can be given by the theory. This reasoning leads also to the crucial idea that probabilities are not defined between individual modalities, but between extravalent classes of modalities; we will come back to this in section 5. 

 \section{Two spin 1/2 particles and Bell's inequalities. }

Another  example with N = 4 is given by two spin 1/2 particles, with spin operators $\vec S_1 = (S_{x1}, S_{y1}, S_{z1})$ and 
$\vec S_2 = (S_{x2}, S_{y2}, S_{z2})$.  Using standard ket notations and $\hbar=1$, the four mutually exclusive modalities when measuring $(S_{z1}, S_{z2})$ are $| m_1= \pm 1/2, m_2= \pm 1/2 \rangle = | \pm, \pm \rangle$. 

But instead of the separated context $(S_{z1}, S_{z2})$, one may use use the joint context $(\vec S^2, S_{z})$, where $\vec S = \vec S_1+\vec S_2$ is the total spin of the two particles. In agreement with $N=4$, and still using standard Dirac notations, one has again  four mutually exclusive modalities, associated with the quantum numbers $S$ and $m_S$, and the kets are related to the previous ones  in the following way :
\begin{eqnarray}
&&| S=1, m_S = +1  \rangle = | +,+ \rangle \nonumber \\
&&| S=1, m_S = -1  \rangle = | -,- \rangle \nonumber \\
&&| S=1, m_S = 0  \rangle = (| +,- \rangle + | -,+ \rangle)/\sqrt{2} \nonumber \\
&&| S=0, m_S = 0  \rangle = (| +,- \rangle - | -,+ \rangle)/\sqrt{2}\rangle \nonumber 
\end{eqnarray}
It is clear that the ket $| S=1, m_S = 1  \rangle$ when measuring $({\bf S}^2, S_z)$ is  the same as $| +,+ \rangle$ when measuring $(S_{z1}, S_{z2})$, same for  $| S=1, m_S = -1  \rangle$ and $| -,- \rangle$. However,  the two contexts are clearly different, and therefore the modalities are also different, but the certainty  in one context is fully transferred to  a certainty  in the other context: the modalities are thus extravalent. 

In a Bell scenario, one prepares the two spins in an entangled modality, e.g. the singlet state $| S=0, m_S = 0  \rangle$, and then one moves to  the context $(S_{z1}, S_{z2})$ with separable modalities $ | \pm, \pm \rangle$. Then the results are necessary random, but they are correlated between Alice (with spin $\vec S_1$) and Bob (with spin $\vec S_2$). The usual question is to ask what happens at Bob's remote place, when Alice performs a measurement on her particle. The answer is obviously that nothing happens at Bob's, who is very far away. Nevertheless, from her measurement Alice perfectly knows the modality for Bob's particle, but in her context, and this can only be verified at a later time by bringing together Alice's context and Bob's particle. 
Therefore we do have quantum non-locality - due to the bipartite context+system nature of the modality - but not action or even influence at a distance.   

Such a situation would never occur in classical physics, where each system is believed to carry its own property: it is thus clear that local realism fails, as demonstrated by Bell's theorem, and  it fails both ways: both classical realism and classical locality fail, for the same reason that  the modality belongs to both the context and the system (see  \cite{csm2b}  for more details). 

It is thus possible to conclude  the famous Einstein-Bohr debate \cite{AA}, by writing ``If, without in any way disturbing a system, {\it neither changing the context}, we can predict with certainty (i.e., with probability equal to unity) the value of a physical quantity, then there exists an element of physical reality corresponding to this physical quantity.''  This fits with Einstein's conception of reality, while adding a Bohrian twist: a quantum system must be permanently associated with a classical context.

 \section{Extracontextuality in the Kochen-Specker theorem}
 
 \noindent {\bf A. Bell-Kochen-Specker-Cabello theorem.}

A third way to illustrate our axioms is using  the framework of the Kochen-Specker theorem as presented by Cabello \cite{AC}. A possible measurement set-up, i.e. a context, corresponds to a column in Fig.~2, and for each context there are N=4 mutually exclusive modalities: if one of them is true, or realized, the 3 other ones are wrong, or not realized. 
\begin{figure}[!h]
\centering
\includegraphics[width = 8.5cm]{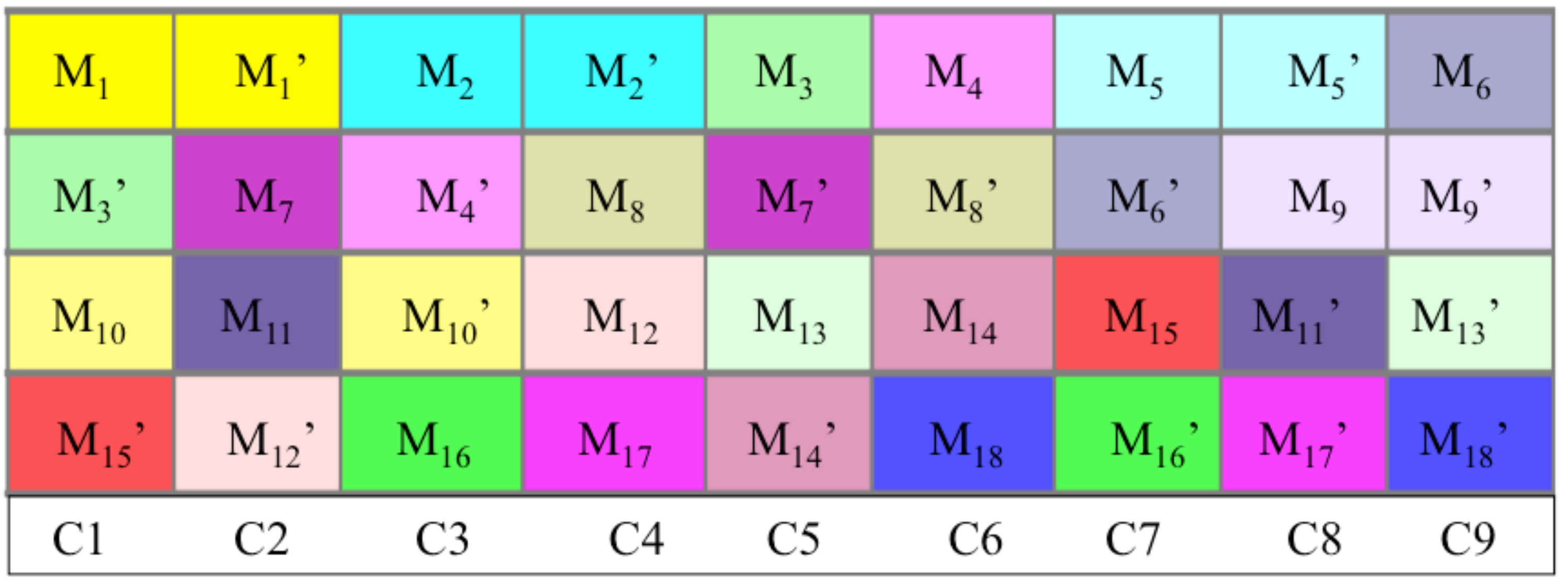}
\caption{Configuration for the Bell-Kochen-Specker-Cabello theorem, with 9 contexts (columns) and 4 mutually exclusive modalities per context. Each of the 18 colored boxes \cite{wiki} associated with modalities is repeated twice, and if a modality is true (realized) in one context, it is automatically true also in the other context where it appears. According to our definitions, the modalities $M_i$ and $M'_i$ (with the same number and color) are not identical, but are extravalent. }\label{change}
\end{figure}

In  Fig.~2 there are 9 different contexts, with the peculiar feature that some modalities (with the same color) are shared between two contexts. This means in practice that if a modality is true in one context, it is automatically true also in the other context where it appears, though the measurement itself may be different;  according to our definitions,  two modalities with the same color are extravalent. We note  that axiom 3 is not fulfilled here, because there is only a  finite number of contexts; we will come back to this later. 

By definition a non-contextual hidden variable theory should obey the following rules : 

1/ one and only one modality can be true for each context

2/ if a modality is true in one context, it must also be true in any other context where it appears (or more correctly according to CSM: if a modality is true in one context, another modality in the same  extravalence class must also be true in any other context where it appears)

3/ any possible modality must be either true (realized) or wrong (not realized). 

Then from 1/ and 3/,  9 slots must be marked true, i.e. one for each context. In addition, from 2/ and 3/, an even number of slots must be marked true.  But  9 is not even, so the three rules cannot hold together. This is the essence of the Kochen-Specker theorem, establishing that there is no non-contextual hidden variable theory.  

On the other hand QM can deal quite well with the situation of Fig.~1, so it must disagree with at least one of the three rules, and it is interesting to ask which one :  

1/ one cannot give up rule 1, because a measurement gives one result among four mutually exclusive ones, this is our second (empirically founded) axiom.

2/ one cannot give up rule 2, because the certainty of a modality can be transferred from one context to another, as required by extracontextuality.  

3/ therefore  one must give up rule 3, i.e. the requirement to give simultaneous truth values to all modalities. This is quite acceptable, since we already know that probabilities are needed, as an unavoidable consequence of our axioms 1 (modalities) and 2 (quantization)  \cite{csm2,csm1}.
\vskip 2mm

 \noindent {\bf B. Extravalence and extracontextuality (again).}

The next question is what should replace the previous rule 3: this is where extravalence comes in. First let rule 2 be restated  in the following way : if a modality is certain in one context, all extravalent modalities are also certain in any other context where they appear. 

But this appears as a particular case of the following more general rule : given a modality in an extravalence class, the probability $p$ to get any other possible modality in another  extravalence class when changing the context depends only on the two extravalence classes, and not on the contexts in which the modalities are embedded. Clearly rule 2 is just a particular case where $p$ = 1, and it can be superseded by this more general rule, which fits with the contextual objectivity of modalities \cite{ph1}: a context is always needed for a modality to show up, but modalities are extracontextual, i.e. their certainty and reproducibility can be transferred between contexts. 

One can thus consider the following rules for extracontextuality:

I /  the $N$ modalities within one context are mutually exclusive, and their probabilities sum to 1: if one is realized ($p$ = 1), all other ones are not ($p$ = 0).

II /  given a modality $u_i$ within a context, the probability to get any other modality $v_j$ within another context depends only on the extravalence classes of these two modalities, and not on their embedding contexts.

It should be obvious that these two rules agree with quantum empirical evidence.
The question we will consider now is whether these rules are strong enough to deduce the formalism of quantum mechanics, which has not been used so far, except as a reference or example. 

\section{Gleason's theorem.}

A crucial feature of the extracontextuality rules given above is that they can be seen as the physical content of the hypotheses needed to derive Gleason's theorem \cite{gleason,csm2a}. Let us remind here a statement of this theorem: 

{\it Let $f(P_i)$ be a function of rank-one projectors $P_i$ in a real or complex Hilbert space with a dimension larger than 2, to the interval [0,1] of real numbers. Let assume that $\sum_i f(P_i) = 1$ for any set $\{ P_i \}$ of mutually orthogonal projectors (i.e. $P_i P_j = P_i  \delta_{ij}$) verifying $\sum_i P_i = \hat 1$. Then there is a unique positive Hermitian operator $\rho$ with unit trace so that $f(P_i) = Tr(\rho P_i)$ for all $P_i$.}

For our purpose, it is useful to consider the particular case where $f(P_i)$ can reach the value 1 for some projector $P_k$. Then $Tr(\rho P_k) = 1$, but this implies that $\rho = P_k$, i.e.  $\rho$ is itself a projector \footnote{$Tr(\rho P_k) = Tr(P_k \rho P_k)$ is the diagonal element of $\rho$ on the basis vector associated with $P_k$. Since it is 1, all other diagonal and non-diagonal elements are 0, and $\rho = P_k$.}.

So if we assume that each extravalent class is bijectively associated with a $N \times N$ Hermitian rank-one projector $P_i$, i.e. $P_i^\dagger = P_i^2 = P_i$, and that  each set of $N$ mutually exclusive modalities is associated to a set of $N$ mutually orthogonal projectors, then one gets

* (from rule II): given an initial modality, the probability $f(P_i)$ to get a final modality associated with $P_i$ is a function of the rank-one projector $P_i$

* (from rule I): for any set $\{ P_i \}$ of mutually orthogonal projectors verifying $\sum_i P_i = \hat 1$, one has $\sum f(P_i) = 1$

These are just the hypotheses needed for the theorem. 
As it is well-known, an essential assumption in Gleason's theorem is  that $f(P_i)$ does not depend on other modalities (or projectors) associated with $P_i$ to set up a context, as required by  rule II. 

This assumption is sometimes called non-contextuality (of probability assignments), but this terminology is misleading with respect to the Kochen-Specker theorem, as it suggests the contradictory claims that QM is both contextual, and non-contextual. Using our terminology it is neither, but it is extracontextual.  In order to clarify this point, let us give some examples: 

* a {\it non-contextual} theory is one where reproducible properties of systems can be defined independently of any context, or equivalently in a single universal context, so that  a system owns its physical properties. This is typically the case of classical mechanics, and it is also deeply embedded in the hypotheses of Bell's theorem. Clearly this fails in quantum mechanics

* a {\it contextual} theory is one where the working framework (in which the results are defined) has to be changed each time the context is changed; in addition, given Bell's theorem,  this change has to be non-local and instantaneous in order to agree with experiments. There are many variants of contextual theories, either with hidden variables (e.g. de Broglie - Bohm theory), or without, but it is clear that some degree of contextuality is required to agree with QM and experiments. 

As said above, QM is none of that: it is not non-contextual because contexts are required to define modalities, but it is not contextual like a hidden-variable theory would be, requiring instantaneous influence or action-at-a-distance.  This is made clear in our approach, by requiring instead that the object carrying properties is a system within a context \cite{csm2b}.  In addition, extracontextuality asserts that  the certainty of modalities can be carried out  between contexts; this has stringent consequences on the way to calculate quantum probabilities. 

\section{The CSM proof: stochastic and unistochastic matrices. }

According to the above section, the extracontextuality rules fit very well with Gleason's theorem, and allow us to recover Born's rule. But clearly the major required hypothesis is that each extravalence class is bijectively associated with a $N \times N$ Hermitian rank-one projector $P_i$; would it be possible to avoid this assumption ? 

A way for doing that has been presented in  \cite{csm2}, and relies on a specific decomposition of the $N \times N$ matrix $\Pi$ giving the probability $p_{j|i} $ to end up in modality $v_j$ of context $\{ v_j \}$, starting from modality $u_i$ of context $\{ u_i \}$. We note that $p_{j|i} $  is not a conditional probability according to the usual (Bayesian) definition, since ``$u_i$ and $v_j$'' is in general not a modality, and cannot be given a truth value 0 or 1. Since all the probabilities starting from a given $u_i$ sum to one, the matrix $\Pi$ is said to be a stochastic matrix. But according to Born's rule, it should be unistochastic  \cite{csm2}, i.e. made from the square moduli of the coefficients of a unitary matrix; this is what we want to show. In order to get this result we demonstrated the following Lemmas \cite{csm2}: 
\vskip 2 mm

\noindent {\bf Lemma 1:} The elements $p_{j|i} $ of a $N \times N$ stochastic matrix can   be written under the general form 
\begin{equation}
p_{j|i} =   \mathbf{Tr}\left( P_i'   \;   R \; P_j''     \;  R \right)
\label{svdthe}
\end{equation}
where $\{ P_i'  \}$ and $\{ P_j''  \}$ are two sets of $N$ hermitian projectors of dimension $N \times N$, mutually orthogonal within each set, and where $R$
is a real nonnegative $N \times N$ diagonal matrix such that $ \mathbf{Tr} ( R^2 )=N$, and $ \mathbf{Tr} (P_i'   \;   R^2 )=1$
for all projectors $P_i' $ within $\{ P_i'  \}$.
\vskip 2 mm

\noindent {\bf Lemma 2:} If  $R=\hat 1$ then the  matrix $\Pi$ is unistochastic (the reciprocal is not true). 
\vskip 2mm

Though $N \times N$ hermitian projectors show up in Lemma~1, there is nothing quantum yet, since this formula applies (in a non unique way) to any given stochastic matrix $\Pi$. In particular, $R$ depends on the particular  $\Pi$ being considered, which is generally not unistochastic. 
The main issue is then to show that $R=\hat 1$ for all pairs of contexts, so that eq.~\ref{svdthe} turns into Born's rule.

This can be obtained by assuming that $p_{j|i}$ takes the specific form: 
\begin{equation}
p_{j|i} =  \mathbf{Tr} ( P_{\tilde u_i}'   \;   R_{\tilde u_i , \tilde v_j} \; P_{\tilde v_j}''     \;   R_{\tilde u_i , \tilde v_j} ).
\label{ecc}
\end{equation}
where $\tilde u_i $ (resp. $\tilde v_j $) is the extravalence class associated with $u_i$ (resp. $v_j$). 
This equation appears as a consequence of rule II, i.e. that given a modality $u_i$ within a context, the probability to get any other modality $v_j$ within another context depends only on the extravalence classes of these two modalities, and not on their embedding contexts. From the parametization of  eq.~\ref{svdthe}, $R$ is fixed for a given pair of contexts containing respectively $u_i$ and $v_j$, and eq.~\ref{ecc} adds the extra constraint that it is also fixed for a given pair of extravalence classes $\tilde u_i$ and $\tilde v_j$, whatever the embedding contexts. 
Then it can be shown (see \cite{csm2} or Appendix below) that $R=\hat 1$ for all pairs of contexts, and Born's rule follows since one has $p_{j|i} =  \mathbf{Tr} ( P_{u_i}' \;  P_{v_j}'')$. 

 It may be useful to compare this reasoning with Gleason's theorem:

* Gleason's theorem first associates each modality $v_j$ with a projector $P_{v_j}$  
acting in the Hilbert space $R^N$ or $C^N$.  Then it looks for $f(P_{v_j})$, with  $\sum_j f(P_{v_j}) = 1$  for any set of $N$ orthogonal  projectors such that   $\sum_j P_{v_j} = \hat 1$, and it concludes that $f(P_{v_j})$ is continuous, and is given by Born's formula $\mathbf{Tr} ( P_{u_i} \, P_{v_j})$, where $P_{u_i}$ is the projector associated with the initial modality $u_i$. 

* The  CSM proof  exploits a generic form of stochastic matrices as a Trace depending on projectors, which is valid generally  even outside QM, as long as $R \neq \hat 1$. Then it uses extracontextuality as expressed by eq. \ref{ecc}, and also continuity of the changes of contexts, in order to get  Born's formula and the quantum formalism. 

So the two approaches start   from different sides, but deal with the same issue of getting an extracontextual probability law, and they eventually come to the same conclusion, i.e. Born's rule. This clearly shows that, though Gleason's theorem is often said to be pure mathematics without physical content, the CSM approach provides it with a well-grounded  physical basis.

\section{Take-home message.}

The main purpose of this article is to provide another view on the abstract reasoning of \cite{csm2},  giving it some flesh by starting from the exemple of idealized thought-experiments. This allows one to better understand the meaning of CSM, and also of extracontexuality and extravalence.
Maybe it is worth again emphasizing that in our approach there is nothing like a ``wave function'', universal or not. The real physical objects are the combinations of a system and a context, and these objects  can be given objective (certain, reproducible) properties called modalities. Contextual quantization  (axiom 2) implies that modalities are related probabilistically between different contexts, and the QM formalism is a mathematical way to calculate these probabilities. 

As a take-home message, the modality is  a real  phenomenon involving a system 
and a context as physical objects, and 
it should be carefully distinguished from the state vector or projector, which is a
mathematical object, associated to a class of extravalent modalities. Mixing up these two notions 
generates a lot of confusion, that may easily be avoided in our approach \footnote{We note that this CSM point of view cannot be easily matched onto the so-called psi-ontology theorems, because the ontological hypotheses (or the objects that are considered) are quite different.}. 

\section{Appendix}

In this appendix we include some missing pieces in the demonstration that $R = \hat 1$; this is a summary adapted from \cite{csm2}. 
As explained above,  we look for stochastic matrices connecting probabilities between any two contexts, because stochasticity is a minimum requirement for consistency of probabilities in our approach. Then we use the general parametrization of lemma~1,
involving two sets of orthogonal projectors and a diagonal  matrix $R$.

If the two contexts are the same, then the stochastic matrix is $\hat 1$, and from its definition $R = \hat 1$ also. Therefore an arbitrary set of mutually orthogonal projectors can be associated with the context, with one (rank 1) projector associated with each modality. We note that the projectors are still undefined at that stage, it is only required that they make an orthogonal set.

Then one goes  from this initial context $C_1$ to another context  $C_2$ with some stochastic matrix, and adjust the initial projectors to fit with the parametrization using the $R$  matrix, which is given for two given contexts. Then one can sum the expression of Lemma~1  over the modalities of $C_2$, and from the normalization conditions of the stochastic matrix $\mathbf{Tr} ( P_{\tilde u_i}'   \;   R_{\tilde u_i , \tilde v_j} ^2 )=1$ and of the projector $\mathbf{Tr} ( P_{\tilde u_i}'  ) =1$ one gets a set of homogeneous linear equations
\begin{equation}
\mathbf{Tr} ( P_{\tilde u_i}'   \;  ( R_{\tilde u_i , \tilde v_j} ^2  - \hat 1)) =0
\label{hecc}
\end{equation}
where the unknowns are the $N$ diagonal coefficients of $(R_{\tilde u_i , \tilde v_j} ^2 - \hat 1)$, and 
the coefficients depend on the set of projectors associated with the extravalence class in the initial context.
If $R_{\tilde u_i , \tilde v_j} ^2 \neq \hat 1$, the determinant of this set of equations must be zero, which amounts to a condition on the set of projectors $\{ P_{\tilde u_n}' \}$ associated with the initial context $\{ u_n \}$.

Now make a small change in the initial context $\{ u_n \}$, keeping the reference modality $u_i$ constant. By continuity, this amounts to changing the projectors except one, and $R$ should not change. But this is clearly impossible: if the projectors change except one (which requires $d \geq 3$), the determinant of eq.~\ref{hecc} will not be zero any more; and even if the changes are restricted so that the determinant remains zero, the set of (degenerate) equations will change, and $R$ cannot remain the same, unless $R = \hat 1$. 
This reasoning can be done for any initial modality and context, so $R = \hat 1$ for any initial context, and thus for any pair of contexts. Therefore the stochastic matrices become unistochastic, which was the crucial point to be demonstrated.

Assuming from now that $R = \hat 1$ between any pair of contexts, the set of initial projectors can be freely chosen (choice of a fiducial basis), and what actually matters are the unitary matrices connecting the sets of mutually orthogonal projectors associated with the different possible contexts. These unitary matrices have to be complex to continuously connect  all permutations to the identity matrix \cite{csm2}. 
For consistency of the construction one also gets that the unitary matrix connecting two contexts in reverse order is the inverse unitary matrix, and thus the ``reverse'' probability matrix is the transposed one, which was not obvious from the start.

Coming back to the thought-experiment of Fig.~1, and considering that the lines are optical fibers, one gets arbitrary change in context by connecting pairs of lines using arbitrary beam-splitters, with suitable  complex  reflection and transmission coefficients. It was shown in \cite{ZZ} that any unitary transformation can be reconstructed in this way. Therefore the ``mapping'' of physical vs. mathematical changes in  the contexts is particularly simple here, but this is not always the case. 

One may also wonder what happens with the scheme of Fig.~2 with only 9 contexts, where Gleason's or CSM theorems cannot be used due to the lack of continuity. The simplest way to solve this issue it to consider that, according to our physically-based axiom 3,  the discrete set of contexts must be embedded in a continuous set; then the previous proofs are applicable again. This follows the CSM idea that the quantum formalism results from the interplay between the quantized number of  modalities accessible to a quantum system, and the continuum of  contexts that are required to define these modalities. 

\vskip 2mm

{\bf Acknowledgements.} 
The authors thank 
Cyril Branciard, Anthony Leverrier, Franck Lalo\"e,
Ad\`an Cabello,  Alastair Abbott, Hippolyte Dourdent for discussions, 
and Nayla Farouki for continuous support.



\end{document}